\documentclass[runningheads]{llncs}

\usepackage[T1]{fontenc}
\usepackage{graphicx}
\usepackage[table]{xcolor}
\usepackage{array}
\usepackage{makecell}
\usepackage{amsmath}
\usepackage{amssymb}
\usepackage[ampersand]{easylist}
\usepackage{enumitem}
\usepackage[sort,compress]{cite}
\usepackage{hyperref}   \usepackage{booktabs}

\newcommand{\ggt}[1]{{\textcolor{black}{{#1}}}}

\makeatletter
\renewcommand\subparagraph{\@startsection{subparagraph}{5}{\parindent}{3.25ex \@plus1ex \@minus .2ex}{-1em}{\normalfont\normalsize\bfseries}}
\makeatother
\usepackage{titlesec}
\titlespacing*{\section}{0pt}{8pt plus 2pt minus 1pt}{4pt}
\titlespacing*{\subsection}{0pt}{6pt plus 2pt minus 1pt}{2pt}
\titlespacing*{\subsubsection}{0pt}{4pt plus 1pt minus 1pt}{1pt}

\hypersetup{pdfborder={0 0 0}}
             
\begin{document}

\title{Do you ``see'' what I ``see''? A Multi-panel Visual Question Answer Dataset for Large Language Model Chart Analysis }
\titlerunning{Do you ``see'' what I ``see''?}

\author{Daniel J. Evans\inst{1}\orcidID{0000-0002-6047-6540} \and
J. P. Naiman\inst{1}\orcidID{0000-0002-9397-6189} \and
JooYoung Seo\inst{1}\orcidID{0000-0002-4064-6012} \and
J. Stephen Downie\inst{1}\orcidID{0000-0001-9784-5090}}

\authorrunning{D. J. Evans et al.}

\institute{School of Information Sciences, University of Illinois, Urbana-Champaign, 61820, USA\\
\email{djevans4@illinois.edu}}

\maketitle
\vspace{-1mm}
\vspace*{-6mm}
\begin{abstract}
Digital libraries increasingly host scientific publications whose figures encode primary research findings, yet making this visual content searchable, accessible, and reusable remains an open challenge. Large language and multimodal models (LLMs/LMMs) provide the possibility to automate many figure parsing tasks such as determining data trends or extracting data from plots. In this work, we consider if these models can reliably support task-oriented visual data analysis when presented with synthetic, but realistic, scientific figures that contain multiple charts. Our Visual Question Answering (VQA) process scaffolds our tasks along a four-level description model from basic conceptual answers to interpretive and analytical understanding. Additionally, we outline our process for synthetically generating our single and multi-panel (charts in which there are multiple panels present) plots and testing the data with various LMMs using quantitative analysis.  In general, we find LMMs become less accurate with increasing figure complexity as the number of panels, number of plot elements, and difficulty of questions increase. Finally, because LMM accuracy is expected to change rapidly, our analysis also considers cost relative to performance as a metric to inform future user decisions. We provide the detailed question-answer pairs used in this analysis, along with the code used to generate random plots, question-answer pairs and detailed bounding-box metadata for chart elements. The code and data are available at  \url{https://github.com/ReadingTimeMachine/LLM_VQA_MultiPanel} and \url{https://huggingface.co/datasets/ReadingTimeMachine/visual\_qa\_multipanel} 
\vspace{-2mm}
\keywords{Visual Question and Answering \and AI and Machine Learning Applications in Digital Libraries}
\end{abstract}

\vspace*{-6mm}
\section{Introduction}
Scientific charts, plots, and diagrams are central to the research housed in digital libraries, and the ability to parse, search, and describe their content is a long-standing challenge in the digital libraries community \cite{kashyap_scientific_2023, naiman_figure_2022}. They grant coherence to large data sets by allowing readers to identify patterns, compare values, and examine particularly noteworthy aspects of a dataset \cite{tufte_visual_1983}. However, for many readers, such as those who are blind or low-vision (BLV), interpreting those figures can be challenging. A recent audit of over 11,000 scientific PDFs found only 2.4\% met basic accessibility criteria \cite{wang_improving_2021}, with figure and chart content among the least accessible elements \cite{kim_exploring_2023, lundgard_accessible_2022}.

Large language models (LLMs) and large multimodal models (LMMs) have opened new avenues for interpreting visual data using natural language. By integrating image recognition with language-based reasoning, these systems are able to analyze and respond to questions about figures. Despite their promise, only recently has testing begun to understand how well these models perform task-oriented visual question analyses in a few-shot or zero-shot setting when faced with multiple paneled images. A growing body of work examines multi-panel chart benchmarks \cite{masry_chartqapro_2025}, \cite{li_chartgalaxy_2025}, while others specifically target reasoning across multiple charts \cite{qiyuanfan_multichartqa-r_2025}. Multi-panel plots are charts in which there are multiple panels present. They are often complex with panels meant to be read in conjunction with each other. Systematic evaluation of cross-panel reasoning, where answering a question requires extracting and comparing information from multiple subplots within a single figure remains limited. This is especially problematic as multi-panel charts comprise $\sim$30--60\% of scientific charts \cite{lee_viziometrics_2018,taschwer2018automatic,lee2015separating}.

To address this gap, we present a preliminary study that evaluates several LMMs on a synthetically generated visual question answering (VQA) dataset designed for chart interpretation on multi-paneled charts. Our dataset scaffolds questions from basic factual identification to interpretive and analytical reasoning. This allows us to measure not only whether models can recognize visual features, but also whether they can generalize to more complex analysis tasks. 

Specifically, we investigate:

\begin{itemize}[noitemsep, topsep=0pt, partopsep=0pt, parsep=0pt]
\item Can zero-shot LLMs reliably parse the multi-panel charts that populate digital library collections, and which question categories (description, comparison, inference) succeed or fail?
\item Can LLMs surface chart content for BLV readers of scholarly publications, such as distributions, comparisons, trends, and what is the cost-accuracy trade-off across model tiers?
\end{itemize}

\subsection{Related Work}
VQA is a research process through which open-ended questions are asked and answered about an image and presented as a dataset \cite{agrawal_vqa_2016}. These questions range in complexity from simple yes/no or multiple choice questions \cite{kahou_figureqa_2018} to more advanced descriptions \cite{ahmed_realcqa_2023,han_chartllama_2023} to summarization \cite{rahman_chartsumm_2023} and reasoning/claim checking \cite{akhtar_chartcheck_2024}. Through the codification of a series of questions and answers, VQA provides a framework to benchmark the accuracy of LMMs to interpret images \cite{cai_vip-llava_2024}. More recent work has begun to address multi-panel chart compositions \cite{li_chartgalaxy_2025, masry_chartqapro_2025, qiyuanfan_multichartqa-r_2025}, with some research including partial support in this space \cite{tang_chartmuseum_2025}.

This codification underpins efforts to generate accessible figure captions \cite{belouadi_detikzify_2024, singh_figura11y_2024, tang_vistext_2023} and frame BLV-relevant queries that are often non-data and contextual \cite{kim_exploring_2023}.

While contextual information is useful, our dataset scaffolds those queries along a four-level figure description model \cite{lundgard_accessible_2022}. This incorporates contextual information alongside other tasks as a description model.

\subsection{Four-Level Figure Description Model}

In what follows, we make use of the four-level figure description model as described by \cite{lundgard_accessible_2022, tang_vistext_2023} in order to quantify the accuracy of several LMMs on figure description tasks. During this study, we focus our efforts on Levels 1--3 of \cite{lundgard_accessible_2022}. We describe our analysis levels in the table below:

\begin{table}[!ht]
\caption{Four-level figure description model adapted from \cite{lundgard_accessible_2022, tang_vistext_2023}.}
\label{tab:levels}
\setlength{\tabcolsep}{3pt}
\scriptsize
\centering
\begin{tabular}{|>{\centering\arraybackslash}p{0.05\textwidth}|p{0.30\textwidth}|p{0.50\textwidth}|}
\hline
\rowcolor{white}
\color{black} Level & \color{black} Description & \color{black} Example \\
\hline
L1 & Low-level ``grammar of graphics'' visual encoding & \textit{``The graph shows a single line plot with x-axis `Bob' (0--100) and y-axis `Linda' ($-$20 to $+$20)''} \\
\hline
L2 & Data trends or ``data facts'' \cite{lundgard_accessible_2022, srinivasan_augmenting_2019, wang_datashot_2020} & \textit{``The median of the y-axis data is 50''} \\
\hline
L3 & Complex trends not evident from statistics, in informal language & \textit{``The blue line appears linear, especially for large X''} \\
\hline
L4 & Requires external context/domain expertise & Not included; left to future work \\
\hline
\end{tabular}

\end{table}
As shown in \autoref{tab:levels}, we focus on Levels 1--3. Sighted readers typically find Level 3 and 4 the most useful descriptions, whereas BLV readers find Level 2 and Level 3 the most useful \cite{lundgard_accessible_2022}.
In our analysis, we test various multi-paneled plot ``types.'' Extending \cite{huang_detection_2024, huang_pixels_2024, lu_mathvista_2023, jcdl2025}, we provide a summary of existing VQA datasets (see \autoref{tab:other_datasets}). Included in this table are the ``type'' of chart and corresponding linked dataset, if available. We found that most VQA datasets incorporate multiple chart types, however, we found that few focus on one specifically, link to their original data, or include bounding boxes for chart elements.

Level 3 questions in our dataset ask about visual trends typically not reducible to a single statistic, for example whether a line appears linear, whether a distribution appears symmetric or skewed, or whether two panels show similar overall trends. Ground-truth answers for L3 questions are derived from the known distribution type used to generate the underlying data.

Additionally, as summarized in \autoref{tab:other_datasets}, few existing datasets include multi-panel figures. Recent work has begun to address this gap \cite{masry_chartqapro_2025, li_chartgalaxy_2025, qiyuanfan_multichartqa-r_2025} by including multi-panel compositions. We also found a series of VQA datasets with partial support for multi-panel plots \cite{lu_mathvista_2023, li_scigraphqa_2023, davila_chart-info_2025, hsu_scicap_2021, tang_chartmuseum_2025}. Yet these datasets primarily collect questions about individual charts within a multi-panel layout rather than posing questions that require reasoning across panels. Furthermore, we found no existing chart VQA dataset that includes contour plots, despite their widespread use in scientific and engineering publications for visualizing continuous surfaces and multi-variable relationships. 

\begin{table}[h!]
\caption{Overview of chart VQA datasets. Plot types: l=line, b=bar, s=scatter, p=pie, h=histogram, o=other. BB: bounding boxes included (P=partial). Data: original data accessible. MP: multi-panel figures included.}
\label{tab:other_datasets}
\centering
\scriptsize
\setlength{\tabcolsep}{3pt}
\renewcommand{\arraystretch}{0.80}
\begin{tabular}{lcccc c}
\toprule
\textbf{Name} & \textbf{Date} & \textbf{Plot Type} & \textbf{BB} & \textbf{Data} & \textbf{MP} \\
\midrule
FigureQA \cite{kahou_figureqa_2018} & 2018 & lbp & $\checkmark$ & -- & -- \\
ChartLlama \cite{han_chartllama_2023} & 2023 & so & -- & -- & -- \\
ChartAssistant/SFT \cite{meng_chartassistant_2024} & 2024 & lbspo & -- & P & -- \\
MMCA \cite{liu_mmc_2024} & 2024 & lbspho & -- & -- & -- \\
Chart2Text \cite{kantharaj_chart--text_2022,obeid_chart--text_2020} & 2020/22 & lbspo & P & $\checkmark$ & -- \\
Chocolate \cite{huang_lvlms_2024} & 2024 & lb & -- & $\checkmark$ & -- \\
Leaf-qa \cite{chaudhry_leaf-qa_2020} & 2020 & lbspo & $\checkmark$ & -- & -- \\
ChartQA \cite{masry_chartqa_2022} & 2022 & lbp & $\checkmark$ & $\checkmark$ & -- \\
ChartCheck \cite{akhtar_chartcheck_2024} & 2023 & lbspo & -- & $\checkmark$ & -- \\
OpenCQA \cite{kantharaj_opencqa_2022} & 2022 & lbspo & $\checkmark$ & P & -- \\
MapQA \cite{chang_mapqa_2022} & 2022 & o & $\checkmark$ & $\checkmark$ & -- \\
MathVista \cite{lu_mathvista_2023} & 2023 & lbspo & -- & -- & P \\
SciGraphQA \cite{li_scigraphqa_2023} & 2023 & o & -- & -- & P \\
VisText \cite{tang_vistext_2023} & 2023 & lbo & P & $\checkmark$ & -- \\
Chart-Info \cite{davila2022icpr,davila_chart-info_2025,davila_icdar_2019,icdar2023,icpr2020} & 2019--24 & lbo & $\checkmark$ & P & P \\
DocVQA \cite{mathew_docvqa_2021} & 2021 & lbp & -- & $\checkmark$ & P \\
ChartFC \cite{akhtar_reading_2023} & 2023 & b & $\checkmark$ & $\checkmark$ & -- \\
RealCQA \cite{ahmed_realcqa_2023} & 2023 & lbso & $\checkmark$ & $\checkmark$ & -- \\
StructChart \cite{xia_structchart_2024} & 2023 & lbpho & -- & $\checkmark$ & -- \\
ChartSumm \cite{rahman_chartsumm_2023} & 2023 & lbp & -- & -- & -- \\
ChartOCR \cite{luo_chartocr_2021} & 2019 & lbp & P & -- & -- \\
LineEX \cite{p_lineex_2023} & 2023 & l & $\checkmark$ & $\checkmark$ & -- \\
SciCap \cite{hsu_scicap_2021} & 2021 & bso & -- & -- & P \\
AutoChart \cite{zhu_autochart_2021} & 2021 & lbs & $\checkmark$ & -- & -- \\
Unichart \cite{masry_unichart_2023} & 2023 & lbp & P & -- & -- \\
PlotQA \cite{methani_plotqa_2020} & 2020 & lbs & $\checkmark$ & $\checkmark$ & -- \\
MatCha \cite{liu_matcha_2023} & 2023 & lbo & -- & $\checkmark$ & -- \\
AI2D \cite{kembhavi_diagram_2016} & 2016 & o & -- & -- & -- \\
TQA \cite{kembhavi_are_2017} & 2017 & o & $\checkmark$ & $\checkmark$ & -- \\
ChartQAPro \cite{masry_chartqapro_2025} & 2025 & lbspo & -- & -- & $\checkmark$ \\
EvoChart-QA \cite{huang_evochart_2025} & 2025 & lbspo & -- & -- & -- \\
ChartMuseum \cite{tang_chartmuseum_2025} & 2025 & lbspo & -- & -- & P \\
ChartGalaxy \cite{li_chartgalaxy_2025} & 2025 & lbspo & -- & $\checkmark$ & $\checkmark$ \\
RefChartQA \cite{vogel_refchartqa_2025} & 2025 & lbp & $\checkmark$ & P & -- \\
ChartDQA \cite{fan_-depth_2025} & 2025 & lbspho & -- & $\checkmark$ & -- \\
ChartQA-X \cite{hegde_chartqa-x_2025} & 2025 & lbp & -- & -- & -- \\
MultiChartQA-R \cite{qiyuanfan_multichartqa-r_2025} & 2025 & lbspo & -- & -- & $\checkmark$ \\
mChartQABench \cite{wei_mchartqa_2026} & 2025 & lbpo & -- & -- & -- \\
What Lies Beneath \cite{naiman_what_2026} & 2026 & h & $\checkmark$ & $\checkmark$ & -- \\
Chart-HQA \cite{chen_chart-hqa_2025} & 2026 & lbpo & -- & -- & -- \\
ChartNet \cite{kondic_chartnet_2026} & 2026 & lbspho & P & $\checkmark$ & -- \\
\bottomrule
\end{tabular}
\end{table}

\section{Methods} \label{section:marking}

\subsection{Figure Dataset Generation}

Synthetically generated data can often fail to mimic realistic images \cite{methani_plotqa_2020,wang_charxiv_2024,masry_chartqapro_2025}, but has the advantage of enabling the storage of precise bounding box information and complex datasets used to create the figures \cite{jcdl2025}. We employ several methods to best align our synthetically generated figures with their features with ``real'' charts as faithfully as possible while enabling the sharing of a dataset which has complete information (dataset, generation parameters) and precise bounding boxes for all chart marks.
Our current pipeline generates four plot ``types'' including line plots, histograms, contour and scatter plots. We generate these synthetic figures using the \textsf{matplotlib} package in \textsf{Python}. Relationships between the x and y data values are chosen from three types of distributions: linear relationships, gaussian mixture models, or random distributions. Colors for lines and bars are chosen at random, however, colors for scatter plot points are chosen from one of the three distribution options. Noise is added to each relationship at a prescribed level. Titles and x and y axis labels are taken from the 1000 most frequent words from a science-based NLP corpus mined from the ar$\chi$iv \cite{naiman2023Ocr}. Inline math formulas are substituted for words randomly, using data from the ar$\chi$iv bulk downloads as in \cite{naiman2023Ocr}. Fonts, fontsizes, scales (log or linear), x/y error bar existence and size, \textsf{matplotlib} plotting style, color maps, DPI, aspect ratio. Plot numbers and layout (horizontal, vertical or rows-columns) are chosen randomly. 

The number of panels in a plot is chosen from a normal distribution centered around 6 with a standard deviation of 15. Each generated plot is automatically checked for overlapping figure boxes or labels, and if found, fontsize and layout is automatically updated until there are no overlaps. We re-run our generation code until there are $>$30\% of figures with multi-panels ($\sim$41\% in this dataset).

\begin{figure}[tbp]
\centering
\includegraphics[width=0.70\textwidth]{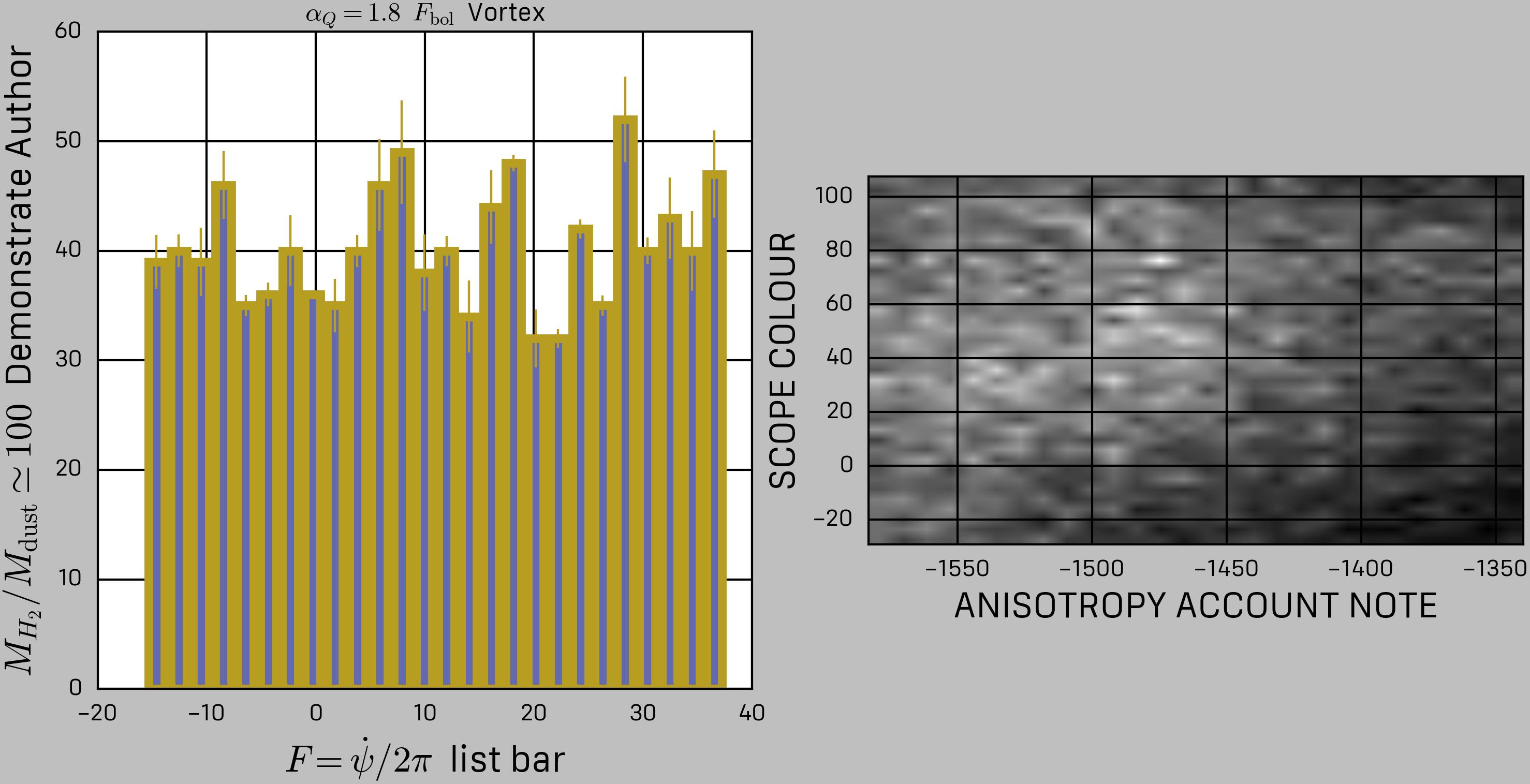}
\caption{Example of a multi-panel figure with complex data marks. On the left is a histogram, while on the right a contour plot is shown.}
\label{fig:complex_figure}
\end{figure}

The randomly selected plot parameters and data are saved in a JSON format, and used to generate VQA pairs using typical LLM/LMM prompting style \cite{sundar_cpapers_2024,al-hadhrami_critical_2023}. While our synthetic figures do not fully capture the visual complexity of real scientific publications, particularly domain-specific color conventions and irregular layouts, they offer complete ground-truth metadata by including features are unavailable in real-figure datasets such as bounding boxes and generation parameters \cite{li2025survey}. Furthermore, as seen in \autoref{tab:other_datasets}, histograms and contour plots are underrepresented in existing datasets but included in ours. All code used to generate these figures is shared on our Github repository.\footnote{\url{https://github.com/ReadingTimeMachine/LLM_VQA_MultiPanel}} Figure images and VQA dataset are stored on HuggingFace.\footnote{\url{https://huggingface.co/datasets/ReadingTimeMachine/visual\_qa\_multipanel}} See \autoref{fig:complex_figure} for an exemplar multi-panel figure.

\subsection{Question Data Generation} \label{section:alignment}
Along with randomly generated figures, our pipeline includes the generation of questions for specific panels in a one to several-panel figure. We restrict questions to those which can be addressed from a model's inspection of the figure alone, without requiring additional contextual information such as article text \cite{huang_lvlms_2024, liu_matcha_2023}, broader documents \cite{masry_unichart_2023}, or domain expertise \cite{lu_mathvista_2023}. We leave this level of analysis (``Level 4'' of \cite{lundgard_accessible_2022}) to future work, focusing instead on what can be extracted from the visual representation itself.

Questions are motivated by the work of \cite{seo_maidr_2024}, with prompts for LLMs distributed into three ``Levels'' of \cite{lundgard_accessible_2022}, either asked of the entire figure (Figure-level type) or of a specific panel (Plot-level type). When asked of a single panel, the panel location is specified with words (e.g., ``second from the top, middle panel''). 

Individual question templates within each Level draw on question types established in prior chart VQA work. These include factual and counting questions comparable to those in ChartQA \cite{masry_chartqa_2022} and PlotQA \cite{methani_plotqa_2020}, trend and comparison questions in the style of VisText \cite{tang_vistext_2023} and OpenCQA \cite{kantharaj_opencqa_2022}, and interpretive reasoning questions adapted from RealCQA \cite{ahmed_realcqa_2023} and MathVista \cite{lu_mathvista_2023}.

Question-answer pairs are generated in four parts and are validated by construction: Level 1 and Level 2 ground-truth answers are computed directly from stored generation parameters (e.g., panel count from figure metadata, means from underlying data arrays), while Level 3 answers derive from the known distribution type used to generate each panel. A manual review of sampled question-answer pairs confirmed correctness. We distinguish Figure-level questions, which require holistic parsing of the full multi-panel layout (e.g., panel counting), from Plot-level questions targeting individual panels. Neither category in the current dataset requires synthesizing values across multiple panels simultaneously. We identify this as the clearest direction for extending the benchmark. Representative question-answer pairs for each level are available in our dataset repository. 

\section{Results and Analysis}

We first pass 200 figures and a single question to each of proprietary LMMs via their API: \textsf{ChatGPT-5.4} \cite{openai_gpt54_2026}, \textsf{Claude-4.5} \cite{anthropic2025claude}, and \textsf{Gemini-3.1} \cite{google_gemini_31_pro_2026}. 
If a token-overflow error is detected, figures are scaled down by a factor of two and the question is re-processed. We restrict our analysis to proprietary LMMs due to the fact that as of writing, open source LMMs generally do not perform as well on chart-specific VQA tasks (e.g., \cite{tang_chartmuseum_2025, han_chartllama_2023}). We consider two ``classes'' of LMMs. Lower level versions of these three models -- GPT-nano with default \textsf{reasoning} (GPT in all figures), Gemini-flash-lite with variable thinking (\textsf{thinking\_budget = -1}, Gemini), and Claude-Haiku (Claude). Mid-to-high level versions of models include GPT-nano with low verbosity but high reasoning (\textsf{verbosity = 'low'}, \textsf{reasoning\_level = 'high'}, GPT-HR), GPT mini with low (high) verbosity (reasoning) (GPT-HRM), and Claude-Sonnet with thinking (\textsf{thinking=\{'type':'adaptive'\}}, \textsf{max\_tokens = 10,000}, Claude-T). These distinctions are not stringently defined, and simply mimic ``buckets'' of potential user parameter choice and model cost.

In order to compare to our pre-generated answers, we parsed LMM results and disregarded answers which did not fit our specifications for output (\ggt{rejecting 130 questions}). For example, we disregard the output if the model failed to return a numerical answer when one was expected. This results in a set of \ggt{33,560 (99.6\%)} answered questions for 200 image files. Each question was submitted as an independent API call with no shared context between questions. This single-query, zero-shot structure is distinct from a model-training setting, where correlated samples from the same figure could bias fitted parameters.

\begin{figure}[tbp]
\centering
\includegraphics[width=0.95\textwidth]{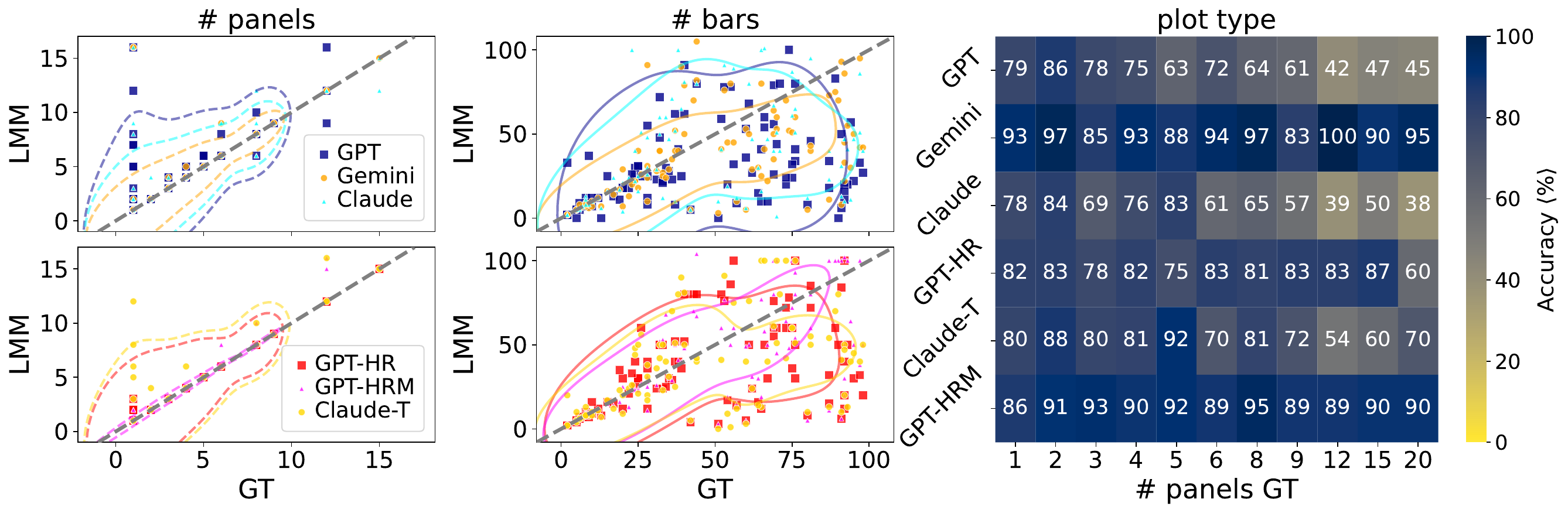}
\caption{Various ``Level 1'' VQA accuracies. Left-most plots show the LMM counts of number of figure panels as a function of ground truth (GT), and center-plots show the LMM counts of number of bars in histogram-type figures as a function of ground truth. Lower-level models (less complex, less reasoning/thinking) are top left/center plots, while mid-to-high level models are on the bottom left/center. Dashed gray line shows where LMM answer is equal to GT, dashed contours show 90\% containment of distribution (left) and solid lines show 50\% containment (center). The right-most plot shows mean accuracy per ``\# panels'' bin of LMM selection of plot type (line, scatter, contour, histogram) as a function of GT ``\# panels''.}
\label{fig:figureLevel}
\end{figure}

The accuracy of LMM answers from ``Level 1'' questions are shown in \autoref{fig:figureLevel}. The left column shows estimates of the number of panels (a Figure-level question), comparing lower-level models (top) with higher-level reasoning models (bottom) as a function of the number of ground truth panels. Dashed lines indicate the 90\% containment region of the distribution. Among the lower-level models, Gemini tracks the ground truth most closely, while GPT tends to slightly overestimate. The higher-level reasoning models (GPT-HR, GPT-HRM, and Claude-T) show equally comparable performance.

The middle column of \autoref{fig:figureLevel} shows estimates of the number of bars for histograms (Plot-level questions), with contours indicating the 50\% containment region. All models perform well at low bars but diverge as the number of bars increase, trending towards underestimation. The higher-level models (bottom-middle plot) show tighter contours, with GPT-HRM performing the best.
The right panel of \autoref{fig:figureLevel} shows the accuracy of each model in identifying plot type (line, contour, scatter, histogram) as a function of the number of panels in the figure. Gemini achieves the highest overall accuracy across panel counts, while GPT and Claude showed a marked decrease in accuracy as the number of panels increased. Accuracy generally remains stable as the number of panels increases, for both Gemini and GPT-HRM.

\begin{figure}[tbp]
\centering
\includegraphics[width=0.99\textwidth]{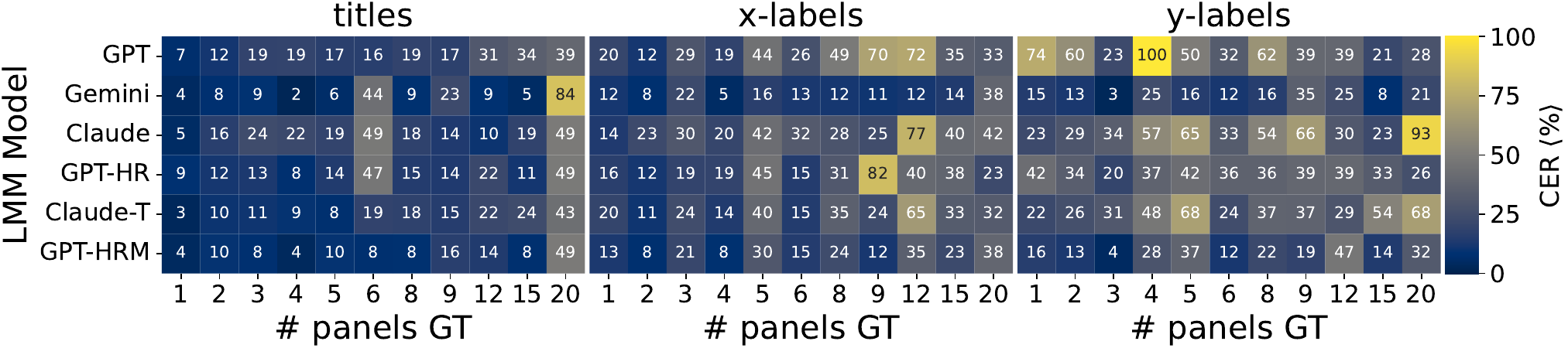}
\caption{Mean character error rate (CER) in each ground truth (GT) ``\# panels'' bin as a function of GT ``\# panels'' for figure panel titles and x/y axis labels. Titles are included with a probability of 50\%, and x/y labels are always present.}
\label{fig:plotLevelElementsNoDiv}
\end{figure}

\autoref{fig:plotLevelElementsNoDiv} shows the mean character error rate (CER) for each model's extraction of panel titles (left), x-axis labels (middle), and y-axis labels (right) as a function of the number of panels. CER was computed by first aligning the ground truth and LMM-returned lists of strings for each figure to minimize overall edit distance. Similar to the right-most panel of \autoref{fig:figureLevel}, mean CER is calculated per ground truth ``\# panels''.  In alignment with OCR and LaTeX recognition work, CER is used as the evaluation metric as labels/titles may include mathematical formulas requiring sensitivity to character-level errors in symbolic notation \cite{gervais2025mathwritingdatasethandwrittenmathematical}.

Across all models, title extraction generally yields lower CER than axis label extraction. This is consistent with the higher relative probability of equations appearing in axis labels as axis labels allow up to 3 words with a 25\% equation probability, compared to 5 words with a 10\% equation probability for titles. 
Y-axis labels show notably higher CER than x-axis labels across most models, likely due to the difficulty of reading vertically oriented text and potential confusion with colorbar labels. Among the models, Gemini achieves the lowest CER overall, with GPT-HRM following closely behind. 

\begin{figure}[tbp]
\centering
\includegraphics[width=0.99\textwidth]{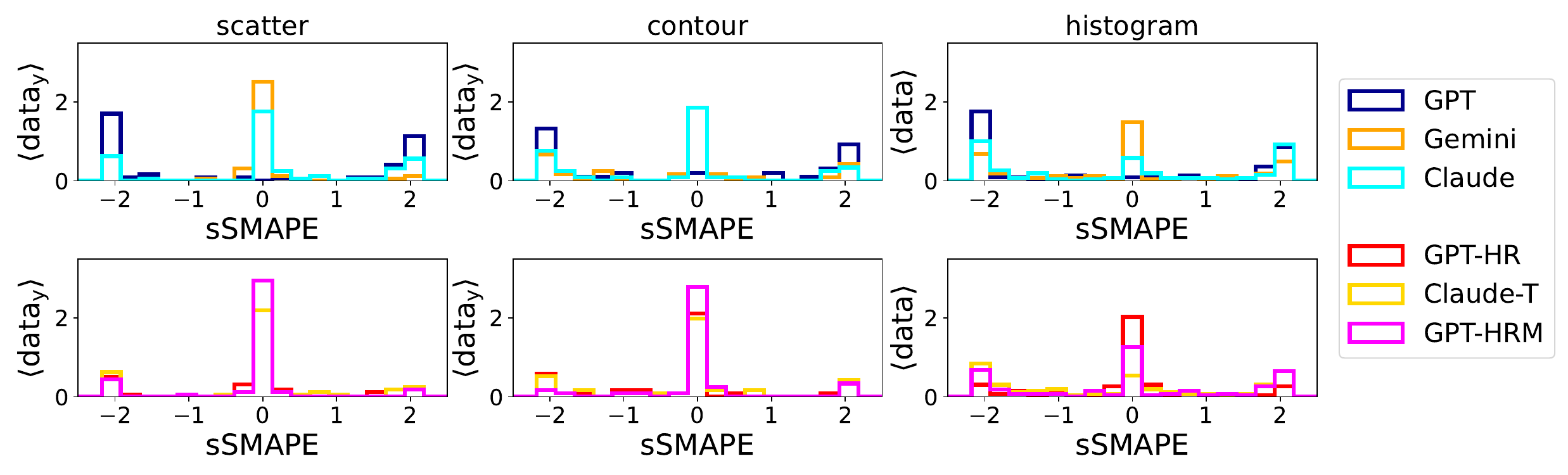}
\caption{LMM accuracy at estimating the mean of the data along the y-axis (scatter, contour) or in the distribution (histogram). Accuracy is measured as $sSMAPE = 2(\mu_{\rm LMM}-\mu_{\rm GT})/(\left\vert \mu_{\rm LMM}\right\vert + \left\vert \mu_{\rm GT} \right\vert)$ \cite{smape1993}. sSMAPE saturation occurs at $\pm$2, where +2 ($-$2) indicates instances where $\mu_{\rm LMM} >> \mu_{\rm GT}$ ($\mu_{\rm LMM} << \mu_{\rm GT}$).}
\label{fig:plotLevel23}
\end{figure}

\autoref{fig:plotLevel23} shows the distribution of normalized error, for Level 2 questions asking models to estimate the mean value data value, broken down by a subset of plot types (contour, histogram, scatter). For scatter and contour plots, the mean data value along the y-axis is requested, while the mean data value (of the distribution) is requested for histograms. Here, the normalized error is calculated as a signed-SMAPE (sSMAPE) where $sSMAPE = 2(\mu_{\rm LMM}-\mu_{\rm GT})/(\left\vert \mu_{\rm LMM}\right\vert + \left\vert \mu_{\rm GT} \right\vert)$ \cite{smape1993}. We use this metric as unlike a simple weighting by $\mu_{\rm GT}$, sSMAPE has the advantage of not diverging when $\mu_{\rm GT/LMM} \sim 0$. Zero-values of sSMAPE occur when $\mu_{\rm LMM} \approx \mu_{\rm GT}$, and saturation of the metric occurs at $\pm$2, where +2 ($-$2) indicates instances where $\mu_{\rm LMM} >> \mu_{\rm GT}$ ($\mu_{\rm LMM} << \mu_{\rm GT}$). Note that for histograms (right panels in \autoref{fig:plotLevel23}), the question asks about the mean of the underlying data rather than the mean along the axis, representing a subtly different, and more difficult, task, resulting in smaller peaks around sSMAPE$\approx$0.

For lower level models (upper panels in \autoref{fig:plotLevel23}), Gemini shows the tightest clustering around zero across plot types while GPT struggles the most with the mean estimate along the y-axis. The mid/higher-level reasoning models (lower panels in \autoref{fig:plotLevel23}) generally improve upon their lower-level counterparts for mean estimation, suggesting that additional reasoning capability helps here.

\subsection{Cost Analysis}
In addition to answer accuracy measurements, there are considerations of cost. 
While ChatGPT-Mini with high reasoning (GPT-HRM) shows overall the highest accuracies, at $\sim$\$55 it also results in the highest spend rate to answer all questions. Claude-Haiku with thinking (Claude-T) is the second most expensive ($\sim$\$36), however it does not meet the overall accuracy of Gemini ($\sim$\$25). 
Other, less expensive models like Claude-Haiku without thinking (Claude) and ChatGPT-nano with high reasoning (GPT-HR) are significantly less than these models ($\sim$\$13 and $\sim$\$11, respectively), however the sacrifice in accuracy is significant.   
At $\lesssim$\$1, ChatGPT-nano (GPT) is the least expensive model, however it is also overall the least accurate. These costs reflect API pricing at time of writing and should be interpreted as relative comparisons rather than absolute benchmarks, as model pricing and performance are subject to rapid change. They are included for methodological transparency.

\section*{Conclusions and Future Work}

This paper presents methods to generate synthetic VQA datasets that can support evaluation of LLM/LMM tools for figure parsing in digital library settings.  
Our work has focused on multi-panel plots, an under represented figure format in the VQA community.
Our error analysis shows that \textsf{ChatGPT}, \textsf{Claude} and \textsf{Gemini} succeed in describing several simple aspects of plots (e.g., number of panels), but that accuracy decreases with increasing figure and question complexity. In other words, LMMs become less accurate as the number of panels, number of plot elements, and difficulty of question increase. These results are broadly consistent with those of single-panel VQA (e.g., Table~6 in  \cite{huang_pixels_2024}), which also show increasing error rates for more complex questions. Our work considers accuracy in combination with cost to evaluate a model's ``precision-per-cost".  While we find Gemini currently provides the overall highest accuracy per cost, it is likely this will change rapidly with future versions of other LMMs.

Our work has used relatively simple metrics (binary comparisons, Levenshtein edit distance) to quantify accuracy.  While this metric works for text like those in our figures, more free-form answers require more sophisticated metrics \cite{liu_mmc_2024,han_chartllama_2023}.  Additionally, recently submitted work highlights the need to baseline any LMM accuracy to that of humans from a range of backgrounds \cite{submittedWork}. 

Because BLV readers rely most heavily on Level 2 and Level 3 content, the accuracy declines we observe on these tasks (\autoref{fig:figureLevel}-\autoref{fig:plotLevel23}) carry direct accessibility implications as the same complexity that challenges LMMs is the complexity most needed to make multi-panel figures usable for BLV readers. The present study focuses on establishing accuracy benchmarks, we leave direct evaluation of model outputs with BLV users for future work. While this work has focused on the estimations of the accuracy of ``zero-shot" LLMs/LMMs, future work additionally will focus on the training of models using the synthetically generated plots discussed in this paper.  As prior work has shown some promise in training and more complex prompting \cite{brown_language_2020}, we expect such methods to decrease the error rates reported here. However, validating this benchmark accuracy against a sample of real multi-panel figures from scientific publications also remains for follow-on work.

\begin{credits}
\subsubsection{\ackname}
This work is supported by a NASA Astrophysics Data Analysis Program Grant (20-ADAP20-0225).

\subsubsection{\discintname}
The authors have no competing interests to declare that are relevant to the content of this article.
\end{credits}

\bibliographystyle{splncs04}  \bibliography{references} 

\end{document}